\documentclass[11pt]{article}
\pdfoutput=1
\usepackage{latexsym}
\usepackage{amssymb}
\usepackage{amsmath}
\usepackage{hyperref, cite, setspace, comment}
\usepackage{graphicx}% Include figure files

\usepackage[height=22cm, width=15.5cm, centering]{geometry}
\usepackage{color}

\begin{document}
\pagestyle{empty}

\begin{flushright}
KEK-TH-2065\\
KEK-Cosmo-226
\end{flushright}

\vspace{3cm}

\begin{center}

{\bf\LARGE  
Unitarity constraint on the K\"ahler curvature
}
\\

\vspace*{1.5cm}
{\large 
Yohei Ema$^a$, Ryuichiro Kitano$^{a,b}$, and Takahiro Terada$^{a}$
} \\
\vspace*{0.5cm}

{\it 
$^a$KEK Theory Center, Tsukuba 305-0801,
Japan\\
$^b$Graduate University for Advanced Studies (Sokendai), Tsukuba
305-0801, Japan
}

\end{center}

\vspace*{1.0cm}

\begin{abstract}
\begin{spacing}{1.2}
{\normalsize

In supersymmetric theories, the signs of quartic terms in the K\"{a}hler
potential control the stability of non-supersymmetric field
configurations.
In particular, in supersymmetric inflation models, the signs are
important for the stability of an inflationary trajectory as well as for
the prediction of the spectral index.
In this paper, we clarify what properties of a UV theory determine the
 sign from unitarity arguments of scattering amplitudes.
As non-trivial examples, we discuss the sign of a four-meson term in
large $N$ supersymmetric gauge theories and also those of the quartic
terms obtained in the intersecting D-brane models in superstring theory.
The UV origins of inflationary models and supersymmetry breaking
 models are constrained by this discussion.

}
\end{spacing}
\end{abstract} 

%%%%%%%%%%%%%%%%%%%%%%%%%%%%%%%%%%%%%%%%%%%%%%%%%%%%%%%%%%%%%%%%%%%%%%%%%%%%
\newpage
\baselineskip=18pt
\setcounter{page}{2}
\pagestyle{plain}
\baselineskip=18pt
\pagestyle{plain}

\setcounter{footnote}{0}

\section{Introduction}

Effective theories are quite strong tools to study low energy
physics especially when the UV theories are strongly coupled or even
unknown, but they usually involve an infinite number of parameters, even
after requiring the symmetry properties.
The analyticity and unitarity of the scattering amplitudes provide a set
of non-trivial (in)equalities for the parameters in the effective
theories~\cite{Martin:1962rt, Pham:1985cr, Ananthanarayan:1994hf, Adams:2006sv,
Jenkins:2006ia,Adams:2008hp}.
After the recent revival of this discussion by Ref.~\cite{Adams:2006sv},
various non-trivial results have been obtained. Examples include the UV
inconsistency of the brane gravity model (the DGP
model~\cite{Dvali:2000hr})~\cite{Adams:2006sv}, the bound on the value
of the superpotential in supersymmetric theories~\cite{Dine:2009sw}, and
the proof of the $a$-theorem in four
dimensions~\cite{Komargodski:2011vj}.

In ${\cal N}=1$ supersymmetric theories, while the superpotential in the
low energy effective theory is strongly constrained by holomorphy, the
K{\" a}hler potential is usually not determined. The quartic terms in
the K\"ahler potential are often important as they modify the shape of
the scalar potential.
For example, in the O'Raifeartaigh models of supersymmetry breaking~\cite{ORaifeartaigh:1975nky}, the
pseudo-modulus field, which is the scalar component of the supersymmetry
breaking chiral superfield, has no potential at the classical level, and
the quartic K\"ahler term generated at the quantum level stabilizes or
destabilizes the vacuum depending on the sign of the coefficient. (See,
{\rm e.g.,} Refs.~\cite{Ray:2006wk, Kitano:2008tm, Komargodski:2009jf} for
discussions of some general features of the O'Raifeataigh models.)

The importance of the quartic terms is not limited to the vacuum of the
theory.  They are relevant whenever supersymmetry is broken,
particularly during inflation.  This is because the quartic terms 
generate supersymmetry-breaking mass terms (Hubble-induced mass terms) of scalar fields in the theory through the coupling to the field whose $F$-term drives inflation.  They
affect the curvature along and/or orthogonal to the inflaton
trajectory, and hence the spectral index and/or the stability
of the trajectory.

In this paper, we study constraints on the sign of the quartic K\"ahler
term from the unitarity arguments of the scattering amplitudes.
We first explain phenomenological motivations in Section~\ref{sec:motivation}.
Then we compare the effective field theory having the quartic K\"{a}hler term with its several UV completions in Section~\ref{sec:EFTandUV}.
Unlike the $\mathcal{O}(p^4)$ terms in the chiral Lagrangian, there are no
universal predictions on the sign of the coefficients~\cite{Low:2009di, Falkowski:2012vh, Bellazzini:2014waa}. The amplitude
analysis in Section~\ref{sec:amp}, however, makes it clear which UV information determines the
sign.  
In Section~\ref{sec:SQCD}, as a non-trivial case with the strong coupling, we examine the supersymmetric QCD at large $N$ and find that the sign is fixed in the leading order of the $1/N$ expansion. 
Since the open-string-like features of the large $N$ expansion play crucial roles, we expect that the same sign is realized in a wide universality class of UV completions, such as the D-brane models in string theory.
This is discussed in Section~\ref{sec:string}.
Finally, Section~\ref{sec:summary} is devoted to a summary and discussion.

%%%%%%%%%%%%%%%%%%%%%%%%%%%%%%%%%%%%%%%%%%%%%%%
\section{Phenomenological motivations}
\label{sec:motivation}
%%%%%%%%%%%%%%%%%%%%%%%%%%%%%%%%%%%%%%%%%%%%%%%

The quartic term in the K\"ahler potential modifies the shape of the
scalar potential when supersymmetry is spontaneously broken.  Thus its
sign is phenomenologically important, e.g., for the stability of the supersymmetry breaking
vacuum or for the inflationary dynamics.

To be specific, we consider the model with
the following K{\"a}hler potential:
\begin{align}
 K = X^\dagger X - {c \over 4\Lambda^2} (X^\dagger X)^2
 + \cdots,
\label{eq:kahler_motiv}
&
\end{align}
where $X$ is a chiral superfield, $c$ is a dimensionless constant, and $\Lambda$ is the UV cutoff scale. 
This form of the K\"ahler potential is general when $X$ is charged under some (global) symmetry, which is the case for all the examples we consider in this paper.
Note that $R$-symmetry is a good symmetry in a wide class of supersymmetry 
breaking models~\cite{Nelson:1993nf}.

The sign of $c$ is important, for instance, when $X$ has a linear superpotential
term, $W = m^2 X$. In this case, the model describes the $F$-term
supersymmetry breaking by $F_X = m^2$. When $c>0$, the minimum is
stabilized at $X = 0$, otherwise one cannot establish the presence of
the vacuum. 
If Eq.~\eqref{eq:kahler_motiv} is a low-energy effective theory of some
UV physics, such as strongly coupled gauge theories or string theories,
it is a non-trivial question whether one can obtain $c>0$ as desired. An
example is SU($N$) supersymmetric QCD with $N_f = N$ flavors of light
quarks discussed in Ref.~\cite{Intriligator:2006dd}, where $c>0$ is
anticipated in Ref.~\cite{Intriligator:2006dd} while no evidence has
been provided~\cite{Katz:2007gv}.

In the context of inflation, 
the sign of the quartic K\"ahler term is important when it contains fields
whose $F$-terms drive inflation, since such a term
generates Hubble induced mass terms.
It controls the curvature of the scalar manifold in two ways: 
(i) the curvature along the inflaton trajectory, 
which is related to the spectral index $n_{\text{s}}$, 
and (ii) the stability of the orthogonal directions to the inflaton trajectory, are affected.
When the trajectory has a bend, the quartic term is related also to
 the sound speed and hence non-Gaussianity~\cite{Hetz:2016ics}.

For instance, the supersymmetric hybrid inflation in its original version~\cite{Dvali:1994ms}
predicts $n_\text{s} \gtrsim 0.98$
which is in tension with the latest Planck observation~\cite{Akrami:2018odb}. 
Once we introduce a quartic K\"ahler term, however, the model
\begin{align}
K =& X^\dag X + \phi^\dag \phi + \overline{\phi}^\dag \overline{\phi} - \frac{c}{4 \Lambda^2} (X^\dag X)^2 + \dots, \label{eq:hybrid_K} \\
W=& \kappa X ( \phi \overline{\phi} - \mu^2 ),
\label{eq:hybrid_W}
\end{align}
gives a lower value of $n_\text{s}$ for a negative $c$~\cite{BasteroGil:2006cm},
while a positive $c$ makes the fit worse,
where $X$ is a gauge singlet (inflaton), and $\phi$ ($\overline{\phi}$)
is a (anti-)fundamental representation of SU($N$) (waterfall field).
A similar situation occurs for 
supersymmetric new (hilltop) inflation~\cite{Kumekawa:1994gx, Izawa:1996dv}, as the model
\begin{align}
K =& X^\dag X - \frac{c}{4 \Lambda^2} (X^\dag X)^2 , \\
W =& v^2 X - \frac{g}{n+1} X^{n+1},
\end{align}
with a small negative $c$ improves the fit to the Planck result 
(unless $n$ is too large)~\cite{Izawa:2003mc}.

As an example of the case~(ii), let us consider 
a chaotic inflation model with a shift symmetry in K\"ahler potential~\cite{Kawasaki:2000yn}:
\begin{align}
K = & \frac{1}{2} (\phi + \phi^\dag)^2  + X^\dag X - \frac{c}{4 \Lambda^2} (X^\dag X)^2 + \dots, \\
W=& X f(\phi),
\end{align} 
where (the imaginary part of) $\phi$ is the inflaton and $X$
is the so-called stabilizer field whose $F$-term drives inflation.
In this class of models, $c$ must be non-negative to ensure the positive mass squared of $X$ during inflation with a generic initial condition.
In models without the stabilizer field~\cite{AlvarezGaume:2010rt, AlvarezGaume:2011xv, Achucarro:2012hg}, the shift-symmetric inflaton potential becomes unbounded from below by the supergravity term 
$-3 e^K |W|^2$~\cite{Kawasaki:2000yn, Achucarro:2012hg}. 
It is cured by introducing the shift-symmetric quartic term~\cite{Izawa:2007qa, Ketov:2014qha,
Ketov:2014hya, Ketov:2016gej}, which also requires $c > 0$.

More generally, we can consider couplings $ -
\frac{c_\psi}{\Lambda^2}X^\dag X \, \psi^\dag \psi$ in the K\"ahler
potential where $X$ is the field whose $F$-term drives inflation and
$\psi$ is some other field.  If $c_\psi$ is positive (and sizable), the
scalar components of $\psi$ obtain a positive Hubble-induced mass
squared so that they tend to be stable during inflation without
generating any isocurvature perturbations.

Finally, we comment on a class of inflation models utilizing the K\"{a}hler potential of the form $K=-3\alpha \log (\Phi +\Phi^\dag)$, such as $\alpha$-attractor models~\cite{Kallosh:2013yoa, Roest:2015qya, Linde:2015uga}.  The quartic K\"ahler term is related to the curvature of the K\"ahler manifold as 
$R = c/\Lambda^2$ at the origin, and the K\"ahler curvature is negative in $\alpha$-attractor models of inflation, $R= -2/(3\alpha) <0$.\footnote{
Indeed, we can rewrite the K\"ahler potential as $K= -3 \alpha \log (1 - X^\dag X)$ under the field redefinition $\Phi = (1+X)/(1-X)$ up to a K\"ahler transformation. In this basis, after expanding it around the origin and taking the canonical normalization, it is transparent that the K\"{a}hler curvature is determined by the quartic term. 
} Similarly, the K\"ahler curvature is negative for inflationary attractors other than the $\alpha$-attractor~\cite{Nakayama:2016eqv}. The role of the quartic K\"ahler term has been discussed in Ref.~\cite{Covi:2008cn} in the context of modular inflation.

%%%%%%%%%%%%%%%%%%%%%%%%%%%%%%%%%%%%%%%%%%%%%%%
\section{UV theories and their effective theories}
\label{sec:EFTandUV}
%%%%%%%%%%%%%%%%%%%%%%%%%%%%%%%%%%%%%%%%%%%%%%%

Before the discussion of the scattering amplitudes, let us consider some
examples that reduce to the model in Eq.~\eqref{eq:kahler_motiv}. The
first model we consider is
\begin{align}
 K& = X^\dagger X + Y^\dagger Y + Z^\dagger Z,
\label{eq:model1K}
\end{align}
\begin{align}
 W& = {\lambda \over 2} X^2 Y + {m } YZ.
\label{eq:model1W}
\end{align}
By integrating out the $Y$ and $Z$ field at tree level, one obtains
\begin{align}
 K_{\rm eff}& = X^\dagger X + {| \lambda |^2 \over 4 m^2} (X^\dagger
 X)^2 + \cdots,
\end{align}
\begin{align}
 W_{\rm eff} = & 0.
\end{align}
We obtain $c < 0$.

The second model we consider is the same K\"ahler potential but with
\begin{align}
 W& = {\lambda \over 2} X Y^2 + m Y Z.  \label{eq:model2W}
\end{align}
This is the O'Raifeartaigh model when we add the linear term of $X$ in
the superpotential. It is known that the model has a minimum at $X=0$ by
evaluating one-loop effective potentials. Indeed, by integrating out $Y$
and $Z$ at the one-loop level, the effective theory is given by
\begin{align}
 K_{\rm eff}& = X^\dagger X - {|\lambda|^4 \over 64 \pi^2 m^2} (X^\dagger
 X)^2 + \cdots,
\end{align}
\begin{align}
 W_{\rm eff} = 0.&
\end{align}
One finds $c > 0$.

We have already seen that there is no universal constraint on the sign of
$c$. There is also an example to give $c>0$ at tree level:
\begin{align}
 K& = X^\dagger e^{-2V} X + m^2 V^2,
\label{eq:vector-modelK}
\end{align}
where $V$ is a massive vector superfield which has the kinetic term,
\begin{align}
 f& = {1 \over 4 g^2} W^\alpha W_\alpha.
\label{eq:vector-modelf}
\end{align}
By integrating out $V$, one obtains the effective
theory: 
\begin{align}
 K_{\rm eff}& = X^\dagger X - {g^2 \over m^2} (X^\dagger X)^2 + \cdots,
\end{align}
\begin{align}
 W_{\rm eff}& = 0.
\end{align}

In the following section, we examine which UV information determines the sign of $c$
by means of the analysis of the scattering amplitude.

%%%%%%%%%%%%%%%%%%%%%%%%%%%%%%%%%%%%%%%%%
\section{Amplitude analysis}
\label{sec:amp}

We consider the scattering amplitude of the scalar component of $X$.
The amplitude of $XX^\dagger \to XX^\dagger$ at a low energy is given by
\begin{align}
 A_{XX^\dagger} (s,t)& = {c \over \Lambda^2} (s + t) + \cdots.
\label{eq:amplow}
\end{align}
The coefficient of the leading term is matched to the one in the
effective theory in Eq.~\eqref{eq:kahler_motiv}, and ``$\cdots$'' represents
higher order terms in the energy expansions.
Since we are interested in the sign of the contact four-point term
obtained by integrating out the UV modes rather than in the loop
contributions of light modes, the latter contribution (if any) is
subtracted in Eq.~\eqref{eq:amplow}.

\begin{figure}[t]
\begin{center}
\includegraphics[width=7cm]{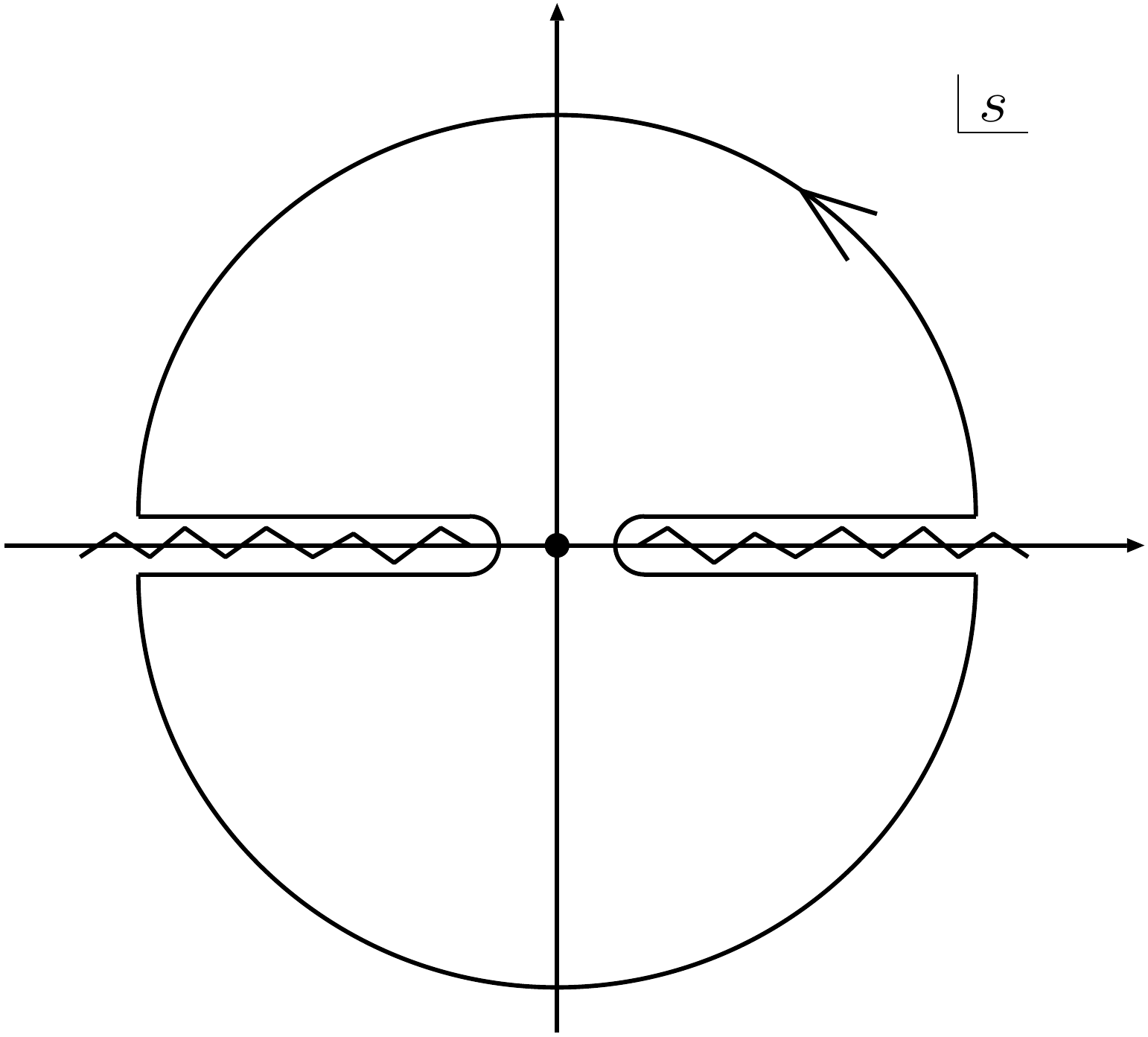} 
\caption{Contour $C$ of the integral.}
\label{fig:contour}
\end{center}
\end{figure}

Following the discussion in Ref.~\cite{Adams:2006sv}, we perform a
contour integral,
\begin{align}
 {1 \over 2 \pi i}&\int_C ds {A_{X X^\dagger} (s,0) \over s^2},
 \label{eq:circle}
\end{align}
in Fig.~\ref{fig:contour}. 
There is a simple pole at $s = 0$ due to Eq.~\eqref{eq:amplow}.
The cuts along the positive and negative real
axes represent the on-shell intermediate states in the
$s$- and $u$-channels, respectively.
If $X$ itself and/or other massless particles contribute to the
intermediate states, one should introduce IR regulators in
Eq.~\eqref{eq:circle} so that the contour enclosing the pole can evade
the cuts.  This IR deformation does not modify the
conclusions~\cite{Adams:2006sv}.

The discontinuity across the cut and/or pole for $s>0$ is given by
\begin{align}
 {\rm Disc}[ A_{XX^\dagger} (s,0)]|_{s > 0}& = 2 i s \sigma_{XX^\dagger} (s),
\end{align}
by the optical theorem, where $\sigma_{XX^\dagger} (s)$ is the total cross
section of $XX^\dagger \to {\rm anything}$, which is positive. By the crossing symmetry, one can also obtain,
\begin{align}
 {\rm Disc}[ A_{XX^\dagger} (s,0)]|_{s < 0}& = 2 i s \sigma_{XX} (-s),
\end{align}
where $\sigma_{XX} (s)$ is the total cross section of
$XX \to {\rm anything}$.

When the large $\vert s\vert$ part does not contribute to the integral, one obtains a relation,
\begin{align}
{ c \over \Lambda^2 }
& = 
{1 \over \pi} 
\int_{0}^\infty ds
\left(
{\sigma_{XX^\dagger} (s) \over s}
-
{\sigma_{XX} (s) \over s}
\right).
\label{eq:relation}
\end{align}
 One can see that $c$ can have both signs in
principle. Note that the sign is determined by the relative sizes between the total cross sections of
$XX^\dagger$ and $XX$ channels.
A similar discussion for four-fermion operators has been given in Ref.~\cite{Adams:2008hp}.

By looking at the examples in the previous section, one can confirm the
above relation. 
In the first example [Eqs.~\eqref{eq:model1K} and \eqref{eq:model1W}],
the $XX \to Z^\dagger$ process is possible at the leading order in
$\lambda$ while the amplitude for the $XX^\dagger \to Y Y^\dagger$
process is $\mathcal{O}(\lambda^2)$, hence $c < 0$. In the second
example [Eqs.~\eqref{eq:model1K} and \eqref{eq:model2W}], only the $XX^\dagger \to YY^\dagger$ or $ZZ^\dag$ processes are possible at
tree level, leading to $c > 0$.  

The third example [Eqs.~\eqref{eq:vector-modelK} and \eqref{eq:vector-modelf}] is a bit tricky. It looks the sign, $c>0$, is
consistent with the fact that only the $XX^\dagger \to V$ contribution is non-vanishing
at $\mathcal{O}(g)$, but actually the large $\vert s\vert$ part of the integral does not
vanish. The contribution from the $t$-channel exchange of $V$ grows
linearly in $s$.\footnote{This is barely consistent with perturbative
unitarity.  The partial wave amplitudes grow logarithmically
corresponding to the fact that perturbation finally breaks down when
logarithms become large.}
 In this case, we cannot use Eq.~\eqref{eq:relation} for the determination of the sign
of $c$.

The relation in Eq.~\eqref{eq:relation} helps us to find out the sign of
$c$ {\em without computing the Feynman diagrams}.
In generic situations, one finds $c>0$. The processes $XX^\dagger \to
{\rm anything}$ are typically possible at tree level, whereas $XX \to
{\rm anything}$ requires an appropriate final state and interactions to
be present.
To conclude it, however, one has to ensure that 
the large $\vert s\vert$ part of the integral is negligible.

%%%%%%%%%%%%%%%%%%%%%%%%%%%%%%%%%%%%%%%%%%%%%%%
\section{Large $N$ supersymmetric QCD}
\label{sec:SQCD}
%%%%%%%%%%%%%%%%%%%%%%%%%%%%%%%%%%%%%%%%%%%%%%%

In the examples so far, we can readily derive the quartic term after integrating out heavy degrees of freedom in the UV models.  In this section, we consider a more non-trivial example: the large $N$ supersymmetric QCD with $N_f$ flavors of light quarks.
In supersymmetric gauge theories, the low energy physics is often described by gauge singlet chiral superfields such as mesons, $M$~\cite{Seiberg:1994bz}. The K{\" a}hler potential for those fields is by
assumption smooth at the origin, and thus can be expanded around it.
For a large $N$, the mesons are weakly coupled, and the scattering amplitudes are assumed to be systematically expanded in powers of $1/N$.

Our main interest is in the situation where 
effective superpotential is non-singular at the origin,
such as the cases with
the quantum deformed moduli space or the s-confinement. That requires
$N_f \geq N$ for the supersymmetric QCD~\cite{Seiberg:1994bz}.
In such a situation, it is more suitable to consider the topological
expansion: taking large $N$ while $N_f/N$ fixed~\cite{Veneziano:1976wm}. 
The s-confinement with a fixed number of $N_f$ is possible once we
extend the theory, e.g.~by including antisymmetric tensor
fields~\cite{Csaki:1996zb} which are treated as double lines in the
$1/N$ expansion.
We here assume that the $1/N$ expansion or the topological expansion
provides a qualitatively correct picture for the scattering amplitudes
of the mesons in the background of $M=0$.
Under the assumption, we will see below that $c>0$ at the
leading order.

At the leading order of the $1/N$ expansion, the low energy effective theory in the meson sector 
is given by
\begin{align}
K=& \text{Tr}(M^\dag M) - \frac{c}{4\Lambda^2} \text{Tr}(M^\dag M M^\dag
 M ),
 \label{eq:k_meson}
\end{align}
where $c$ is $\mathcal{O}(1/N)$.

\begin{figure}[t]
\begin{center}
\includegraphics[height=4cm]{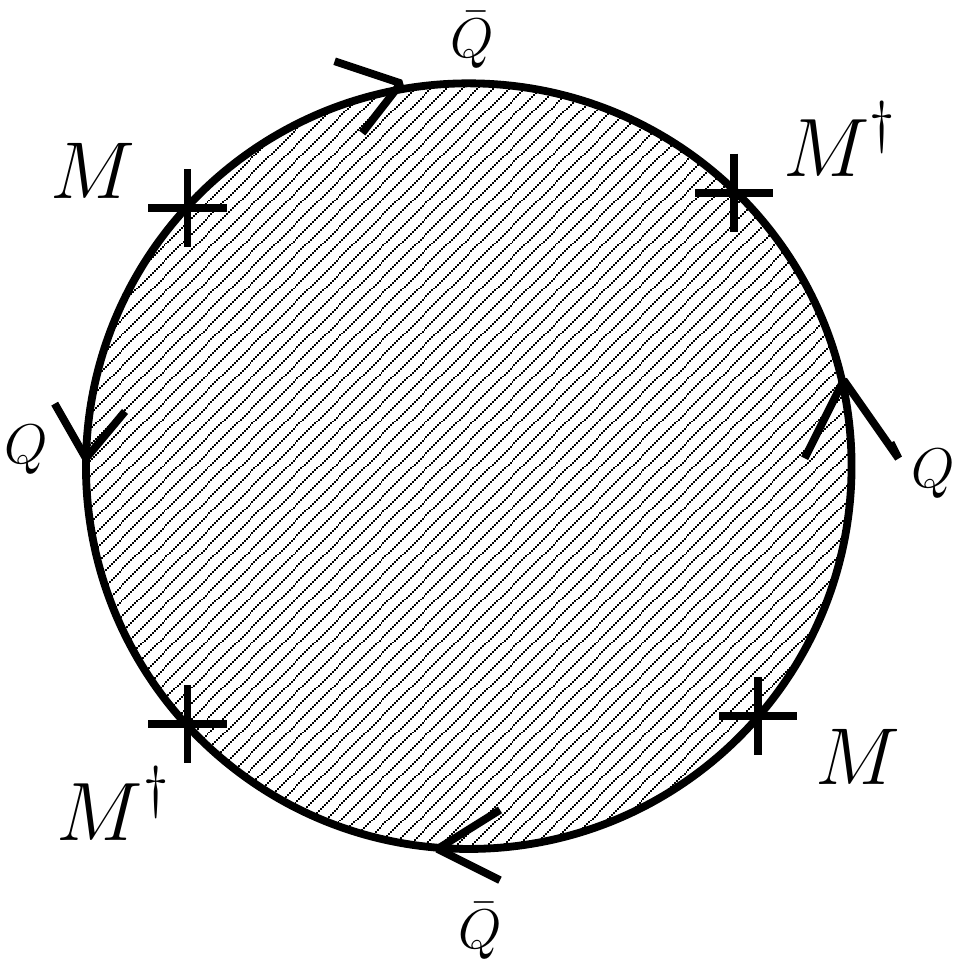} 
\caption{The leading diagram for the meson-meson scattering.}
\label{fig:meson}
\end{center}
\end{figure}

Below, we consider the scattering amplitudes among mesons.
To this end, we decompose the $N_f \times N_f$ meson matrix as
\begin{align}
 M = \sqrt{2} M^a T^a + {1 \over \sqrt N_f }M^0 \cdot {\bf 1},\quad
 (a=1,\cdots, N_f^2-1)
\end{align}
where $T^a$ are the generators of SU($N_f$), so that $M^a$ and $M^0$ are canonically normalized.
For example, the scattering amplitude of 
$M^a M^{0\dagger} \to M^a M^{0\dagger}$ at low energy is
\begin{align}
 A_{MM^\dagger}^{(a0)} (s,t) = 
{c \over \Lambda^2}  (s + t) + \cdots&.
\end{align}
The amplitude at this order is given by the diagram in Fig.~\ref{fig:meson}.  Here,
the arrows represent the flow of chirality, not that of color.
The imaginary part of the forward amplitude, $A_{MM^\dagger}^{(AB)}
(s,0)$, ($A,B=0,1,\cdots, N_f^2-1$), for $s>0$ is the total cross section
of $M^A M^{B\dagger} \to {\rm anything}$.

In the case where $N_f/N$ is fixed, the final states of the $M^A M^{B
\dagger}$ channel include resonances at $\mathcal{O}(1/N)$ as well as
$n$-meson states at $\mathcal{O}(N_f^{n-1}/N^n)$. Therefore, there is a cut in the
real positive axis in the amplitude. For fixed $N_f$, there are only
poles of the resonances.

On the other hand, the final states of the $M^A M^B$ channel do not have
a resonance at $\mathcal{O}(1/N)$. The chirality flow does not allow us to have
three-point vertices. The multi-meson final states are also suppressed by
the same reason.

Therefore, at $\mathcal{O}(1/N)$, the discontinuities of the amplitudes
are only on the $s>0$ axis. By unitarity, they contribute positively to
$c$ as we have seen.
If the large $\vert s\vert$ part of the
integral vanishes, one can conclude $c>0$.
One can separately discuss the large $|s|$ behaviors for the cases of fixed $N_f/N$
and fixed $N_f$.

With $N_f/N$ fixed, the total cross section is dominated by the
multi-particle final states. This is the same situation as the hadron
scatterings in the real QCD where the high-energy scattering is
dominated by the inelastic scattering, i.e.,~by the imaginary part, ${\rm Re} A_{MM^\dagger} (s,0) \ll {\rm Im}
A_{MM^\dagger} (s,0)$.
In such situations, the difference between ${\rm Im} A_{MM^\dagger}
(s,0)/s$ and ${\rm Im} A_{MM} (s,0)/s$ vanishes at a large
$s$ by Pomeranchuk's theorem~\cite{Pomeranchuk:1958,Weinberg:1961zz}. 
Since $\text{Im} A_{M^\dag M} (-s,0)=\text{Im} A_{MM}(s,0) = 0$ for $s \to  \infty$,  $A_{MM^\dag}(s, 0)/s  \to 0$ for $s\to \infty$, which implies that 
 the large $\vert s\vert$ part of the integral vanishes. 
  Therefore, we can conclude $c>0$ at $\mathcal{O}(1/N)$.

With $N_f$ fixed, we may first define $\hat{c}(t)$ as
\begin{align}
	\frac{\hat{c}(t)}{\Lambda^2} \equiv \left.\frac{d}{ds}A_{MM^\dagger}^{(a0)}(s,t)\right\vert_{s=0}
	= \frac{1}{2\pi i}\int_C ds \frac{A_{MM^\dagger}^{(a0)}(s,t)}{s^2},
\end{align}
where the original $c$ is given by $c = \hat{c}(0)$.
There are only poles at the leading order in $1/N$.
The locations of the $s$ poles of $A_{MM^\dagger}^{(a0)}(s,t)$ 
are independent of $t$, and vice versa,
since there is no $u$-channel process at this order.
Thus $A_{MM^\dagger}^{(a0)}(s,t)$ for a fixed $s$ has poles only in the $t > 0$ region,
as there are no tachyons nor massless particles in the $t$-channel.
This indicates that $\hat c(t)$ also has singularities only in the $t>0$
region as one can see from the second expression.

The integral can be decomposed, as usual, to the large $\vert s\vert$
part and the discontinuity part expressed by the imaginary part of the
amplitude.  
By writing the large $|s|$ behavior as the Regge trajectory, $A(s,t) \sim s^{j(t)}$, it has recently been found
that there is a universal behavior, $j(t) \sim \alpha' t + \cdots$, in
the large positive $t$~\cite{Caron-Huot:2016icg}.
For a large negative $t$, which corresponds to the fixed angle
scatterings at high energy, the amplitude should behave according to the
quark counting rule, $j(t) \sim -2$~\cite{Brodsky:1973kr,
Lepage:1980fj}.
Therefore, by taking $t$ at a sufficiently negative value,
the large $\vert s\vert$ part does not contribute.
See Refs.~\cite{Andreev:2004sy} and \cite{Veneziano:2017cks} for
explicit constructions of amplitudes to realize the asymptotic behaviors
at positive and negative $t$.

The function $\hat c (t)$ around $t = t_0$ for a large negative $t_0$ is
expressed as
\begin{align}
 {\hat c (t) \over \Lambda^2}& = {1 \over \pi} \int_0^\infty ds\,
 {{\rm Im} A(s,t) \over s^2}
 = \sum_{k=0}^{\infty} c_k (t_0) (t-t_0)^k,
 \label{eq:xi}
\end{align}
where
\begin{align}
 c_k (t_0)& = {1 \over \pi} \int_0^\infty ds\, \frac{a_k (s,t_0)}{s^2}, 
\end{align}
and
\begin{align}
{\rm Im} A(s,t) & = \sum_{k=0}^\infty a_k(s,t_0) (t-t_0)^k.
\end{align}
The behavior at a large negative $t$ ensures that the integrals converge,
and the expansion around $t=t_0$ in Eq.~\eqref{eq:xi} is defined. Since
the singularities of $\hat c(t)$ are only in the $t>0$ region, the
radius of convergence is larger than $|t_0|$.
In particular, the $t\to 0$ limit of the integral in Eq.~\eqref{eq:xi}
is valid, and by ${\rm Im} A(s,0) > 0$, we conclude $c > 0$.

We stress that the absence of $u$-channel poles is crucial in this
discussion.  If there were $s$ and $u$-channel poles simultaneously,
$\hat{c}(t)$ generally would have poles in the region $t < 0$ since the
locations of the $u$-channel poles are $t$-dependent.
In that case, the radius of convergence could be smaller than $\vert t_0
\vert$. If $A_{XX^\dagger}$ has only $u$-channel discontinuities, we can
instead consider $A_{XX}$ which is related to $A_{XX^\dagger}$ by the
$s$-$u$ crossing.

The above discussion based on the behavior of the fixed angle
scatterings and the location of the singularities of $t$ is not limited
to the case where the amplitudes have only simple poles.
For example, the same discussion could have applied to the case of fixed
$N_f /N$.

The high-energy behavior of the forward amplitude implies that
the Froissart bound~\cite{Froissart:1961ux}, $A(s,0) \lesssim s \log^2 s$, 
is not saturated in both
cases. On the other hand, in the real world QCD, the growth of the total
cross section, $\sigma (s) = {\rm Im}A(s,0) / s$, is experimentally
observed. This may be explained by contributions from higher order terms
in $1/N$, such as the exchange of the
Pomeron. (See, {\rm e.g.,} \cite{Pomeranchuk:1956,Okun:1956,
Foldy:1963zz,Kaidalov:2003au, Ewerz:2003xi}.) 
The subleading contributions, even though they may be more important at
a large $s$, are irrelevant for the current discussion since we are formally
expanding the amplitude in terms of $1/N$, where the unitarity should hold at each order.

Except for gauge boson exchange, it may well be the case that the saturation of the Froissart bound (up to $\log s$) and correspondingly the nonzero value of the large $|s|$ part of the integral are only possible by Pomeron exchange, which implies a universal growth of forward amplitudes irrespective of the quantum numbers.  Assuming the universal asymptotic behavior of the amplitude, the authors of Ref.~\cite{Bellazzini:2014waa} showed that the large $|s|$ contribution can be neglected for the scattering of real representations in theories with global symmetries by exploiting the crossing symmetry.
%The convergence of the dispersion relation in theories with global symmetries has been discussed  the crossing property.

Our discussion seems to support  
 the existence of the meta-stable vacuum in the $N_f = N$ supersymmetric
QCD. In order to establish the existence, however, one also needs to consider the
double-trace terms, $({\rm Tr} (M^\dagger M))^2$, generated at $\mathcal{O}(1/N^2)$ 
from the cylinder diagram as well as meson-baryon terms. The
meson-baryon terms can be ignored once the U($1$) baryon symmetry is
gauged so that the baryons are heavy.
Unfortunately, for the double-trace term, the imaginary part exists for
both channels and also the large $s$ behavior is expected to be such
that the large $\vert s\vert$ part does not vanish. It is difficult in this situation
to determine the sign. 
One may ignore the double-trace term if the effects of the quark loops
are parametrically suppressed so that the $\mathcal{O}(N_f/N)$ terms are
counted as subleading contributions.
This assumption may be supported (although not guaranteed) by the fact
that the theory is in the confining phase rather than in the conformal
window as well as by the success of the OZI rule~\cite{Okubo:1963fa,
Zweig:1964jf, Iizuka:1966fk} in the real QCD with $N_f = N = 3$.
%

%%%%%%%%%%%%%%%%%%%%%%%%%%%%%%%%%%%%%%%%%%%%%%%
\section{String theory amplitudes}
\label{sec:string}
%%%%%%%%%%%%%%%%%%%%%%%%%%%%%%%%%%%%%%%%%%%%%%%

We have discussed that $c$ is positive by considering the disk amplitude
in the large $N$ theory. Because of the similarity between the large $N$
expansion of gauge theories and the $g_{\text{s}}$ expansion of
perturbative string theory, our discussion may be applicable also to the
open-string sector of string theory.  

The quartic K\"ahler terms, which correspond to dimension-six
four-fermion terms, are generated by disk amplitudes of the open strings
in intersecting D-brane backgrounds.
In models with $N$ D$p$-branes and $N'$ D$p'$-branes, U($N$) and U($N'$)
gauge fields arise from strings whose ends are on the D$p$-branes and
D$p'$-branes, respectively.  We also have bifundamental matter $X$ from
strings which end on a D$p$-brane and a D$p'$-brane. Considering the
scattering of the bifundamental matter $XX^\dag \to X X^\dag$, there are
no $u$-channel poles nor cuts at the leading order of perturbation. This
tells us that the amplitude has no imaginary parts along the real $s<0$
line for the forward scattering.  On the other hand, we have an infinite
tower of massive intermediate string states for the $s$-channel.  In
Section~\ref{sec:amp}, we have seen that their contributions are positive when
evaluating $c$.

For example, the four-fermion amplitude for the proton decay in
Ref.~\cite{Klebanov:2003my} can be matched to the quartic K\"ahler term
at low energy. Since the sign is the same as that in the case of the massive gauge boson
exchange, it corresponds to $c>0$.
On the other hand, a similar analysis in Ref.~\cite{Antoniadis:2000jv}
for the contact four-fermion interactions reported that the
contributions to the coefficients from the string oscillator modes turn
out to have the opposite sign to the counterpart of QFT contributions, and
thus $c<0$.

Both amplitudes are given in the form of
\begin{align}
 A_{XX^\dagger} (s,t)& = g_{\text{s}} l_{\text{s}}^2 (s+t) \int_0^1 dx \, x^{-1-l_{\text{s}}^2 s}
 (1-x)^{-1-l_{\text{s}}^2 t} G(x),
\end{align}
where $g_{\text{s}}$ and $l_{\text{s}}$ are, respectively, the string
coupling constant and the string length, and $G(x)$ represents
model-dependent non-negative functions of $x$.  In both of the examples,
one can explicitly see that $A_{XX^\dagger} (s,0)/s$ vanishes at a large
$|s|$.
In this situation, one should find $c>0$ as there is no discontinuity
for $s<0$.
By looking closely at the amplitude which gives $c < 0$, 
we find that the sign of the residues of
the string oscillator poles are opposite to those of gauge bosons,
which seems to violate unitarity.

When $X$ is embedded as above (and if string theory is unitary), $c>0$
is quite general. 
In the actual applications such as for the inflation models, we are
interested in the coefficients obtained by integrating out massive fields which
are absent in the effective theory.
In this case, the amplitude has the same structure as the large $N$
theory with $N_f$ fixed. By the same discussion around Eq.~\eqref{eq:xi},
the large $|s|$ behavior should be good enough for a negative $t$, and one
should be able to take the $t \to 0$ limit smoothly.

%%%%%%%%%%%%%%%%%%%%%%%%%%%%%%%%%%%%%%%%%%%%%%%
\section{Conclusion}
\label{sec:summary}
%%%%%%%%%%%%%%%%%%%%%%%%%%%%%%%%%%%%%%%%%%%%%%%

Unitarity is a fundamental requirement of the quantum theory, and is
incorporated automatically in the definitions of the theory. However,
when we consider the low energy effective theories, the implementation
of the unitarity becomes quite non-trivial. The unitarity looks just
violated at high energy. This actually turns out to be good news. By
requiring that the UV theory is unitary, one can constrain the parameters
in the effective theory.

For dimension-eight operators such as $(\partial_\mu \pi)^4$, the
corresponding scattering amplitudes are $\mathcal{O}(s^2)$. In this case, it
has been shown that the coefficients are positive by
unitarity~\cite{Adams:2006sv}. This is quite surprising when we are
working within the effective theories. Both signs looked allowed in
principle from the low energy perspective, at least in the trivial background.

We extend the study to dimension-six operators such as $|\phi|^2
\partial^\mu \phi^\dagger \partial_\mu \phi$.  In this case, the sign of
the coefficient is not universally determined while it is controlled by
the relative sizes of the total cross sections of $\phi \phi^\dagger$
and $\phi \phi$ scatterings if the high energy behavior is good
enough. 
In contrast to the case with the dimension-eight higher-derivative
terms, the dimension-six term itself never shows the violation of the
global notion of causality such as superluminality in low energy.  When
a particular class of UV theories is assumed, however, the sign of the
dimension-six operator in the effective theory is restricted.  We found
that in supersymmetric gauge theories, the single-trace four-meson
operator in the K\"ahler potential has a negative sign ($c>0$, see
Eq.~\eqref{eq:kahler_motiv}) because there are vector resonances in the
$MM^\dagger$ channel whereas no resonance in the $MM$ one.  The analysis in this paper applies not only for four-scalar operators, but also for
four-fermion operators, $\bar \psi \gamma^\mu P_L \psi \bar \psi
\gamma_\mu P_L \psi$ and boson-fermion operators, $i \bar \psi
\gamma^\mu P_L \psi \phi^\dagger \partial_\mu \phi + {\rm h.c.}$
 A similar discussion was given in Ref.~\cite{Adams:2008hp} for the scattering amplitudes of heavy fermions rather than the massless ones in which we are interested. See Ref.~\cite{Bellazzini:2016xrt} for the extension to the cases with an arbitrary spin. 
There have also been discussions on the sign of anomalous dimensions 
based on unitarity~\cite{Higashijima:2003et}.

The dimension-six terms appear in various kinds of
phenomena. 
For example, in the context of inflation, the negative K\"ahler terms are favored or disfavored depending on
models. This means that the UV completion of each model is restricted in
a certain direction. For example, the hybrid inflation model should not
be UV completed to supersymmetric QCD-like theories or open string
sectors in string theory.

It is sometimes the case that there are no possible intermediate
states for $u$-channel processes, hence no poles nor
cuts along the real negative $s$ for the forward scattering.  In such a
situation, $c>0$ once the large $\vert s\vert$ part of the contour vanishes.  
Similarly, if there are no poles nor cuts on the real
positive $s$ for forward scattering, one can conclude $c<0$ once 
the large $\vert s\vert$ part vanishes.  Therefore, we list up several
situations in which we can establish that the large $\vert s\vert$ part of the
integral vanishes.
With the discontinuities only in either a positive or negative side of the real axis,
the large $\vert s\vert$ integral can be ignored if any one of the following conditions is met:
\begin{itemize}
\item We can directly confirm that the amplitude is damped fast enough at high energy, i.e., $|A(s,0)/s| \to 0$ as $|s|\to \infty$.
\item There is a classical picture of scattering by a geometric
      cross section at high energy so that $\text{Re}\,A(s,0) \ll
      \text{Im}\, A(s,0)$ is
      ensured~\cite{Weinberg:1995mt}.
\item The ratio $A(s,t)/s \to 0$ as $|s| \to
      \infty$ for a negative $t$, and there is no massless 
      (nor tachyonic) state mediating $t$-channel processes.
\end{itemize}
The first case is trivial. We directly confirmed it in the case of the string amplitudes. 
The second criterion is met in QCD-like theories as discussed in the
case of large $N$ theories with $N_f/N$ fixed. 
In the third case, we utilize the convergence in the $t<0$ region and
extrapolate it to $t=0$.  The extrapolation is justified in the absence
of the $t=0$ pole in the scattering amplitude.  This argument was used
in the case of large $N$ theories with $N_f$ fixed.

The quartic term of the K\"ahler potential is essentially the K\"ahler curvature.
Applying our discussion to the cases with the nonvanishing scalar background, $\langle X \rangle \neq 0$, we may infer the geometry of the K\"ahler manifold within the region whose size is at most the cutoff scale, $|X| \lesssim \Lambda$. 
It is interesting to note that parameters of the effective theory are further constrained non-trivially depending on the topology of the K\"ahler manifold.
For example, the parameters of the supergravity models are quantized depending on the topology~\cite{Witten:1982hu}.

Finally we comment on higher dimensional operators.
Although we focused on the dimension-six 
operators such as $\vert \phi \vert^2 \partial^\mu \phi^\dagger \partial_\mu \phi$,
the discussion is readily extended to operators which include higher derivatives.
For instance, let us consider an operator constructed from four scalar fields 
and $2m$ derivatives.
In the low energy limit, the amplitude due to such an operator should behave as $\mathcal{O}(s^m)$,
and hence we can consider
\begin{align}
 {1 \over 2 \pi i}&\int_C ds {A (s,0) \over s^{m+1}},
\end{align}
with the contour in Fig.~\ref{fig:contour}, to examine the sign of the coefficient.
We studied $m=1$ in this paper, and $m=2$ (and also other even $m$) is discussed in Ref.~\cite{Adams:2006sv}.
For $m \geq 2$, the discussion is greatly simplified since the large $\vert s\vert$ part 
always vanishes due to the Froissart bound. 
Thus for $m \geq 2$, the sign is always positive for an even $m$,
while the relative size of $s$- and $u$-channel processes determines the sign for an odd $m$.

\section*{Acknowledgements}
This work is supported by JSPS KAKENHI Grant No.~JP18J00540 (YE),
JP15H03669 (RK), JP15KK0176 (RK), JP17J00731 (TT), MEXT KAKENHI Grant
No.~JP25105011 (RK), JP18H05542 (RK), and JSPS Research Fellowship for Young Scientists
(YE, TT).

\bibliographystyle{utphys_us}
\bibliography{kahler_ref}

\end{document}